# The Pace and Pulse of the Fight against Coronavirus across the US, A Google Trends Approach

**Authors:** Tichakunda Mangono (1), Peter Smittenaar (1), Yael Caplan (1), Vincent Huang (1), Staci Sutermaster (1), Hannah Kemp (1) and Sema K. Sgaier (1,2,3) ((1) Surgo Foundation, Washington DC, 20011, USA, (2) Department of Global Health & Population, Harvard T.H. Chan School of Public Health, Boston MA, 02115, USA, (3) Department of Global Health, University of Washington, Seattle, WA 98104, USA)

**Abstract:** The coronavirus pandemic is impacting our lives at unprecedented speed and scale - including how we eat and work, what we worry about, how much we move, and our ability to earn. Google Trends can be used as a proxy for what people are thinking, needing, and planning. We use it to provide both insights into, and potential indicators of, important changes in information-seeking patterns during pandemics like COVID-19. Key questions we address are: (1) What is the relationship between the coronavirus outbreak and internet searches related to healthcare seeking, government support programs, media sources of different ideologies, planning around social activities, travel, and food, and new coronavirus-specific behaviors and concerns?; (2) How does the popularity of search terms differ across states and regions and can we explain these differences?; (3) Can we find distinct, tangible search patterns across states suggestive of policy gaps to inform pandemic response? (4) Does Google Trends data correlate with and potentially precede real-life events? We suggest strategic shifts for policy makers to improve the precision and effectiveness of non-pharmaceutical interventions (NPIs) and recommend the development of a real-time dashboard as a decision-making tool. Methods used include trend analysis of US search data; geographic analyses of the differences in search popularity across US states during March 1st to April 15th, 2020; and Principal Component Analyses (PCA) to extract search patterns across states.

# Introduction

Over a period of a few months, the COVID-19 pandemic has dramatically and rapidly changed the lives of most Americans. With the spread of coronavirus, people have faced extreme uncertainty due to evolving information, calls for behavior change, and economic shocks. As states move through different stages of the response and plan a path back to normalcy, it is important to develop ways to gather and measure insights into how Americans are responding to the challenges and uncertainties posed by the pandemic.[1] This will help policy-makers know what to focus on as we continue to navigate the COVID-19 outbreak. In particular, is it important to understand the changes and patterns in Americans' information-seeking over the course of the pandemic, both as a window into key concerns as they emerge and as potential signals of actual behavior change.

Collecting data on population responses to a crisis poses a host of methodological challenges.[2,3] Surveys are a commonly used way to gather information on how people are reacting to a health crisis.[4] Several surveys have been deployed to measure Americans' reactions to COVID-19, and they offer insights into the concerns and behaviors of many Americans.[5,6] However, traditional large-scale surveys are expensive and time-consuming, and the rapidly changing nature of the pandemic makes it difficult to capture time-sensitive relevant data.[4] Another option is to use big data, generating insights from existing large-scale health data like electronic patient records and surveillance systems.[7] However, working this type of data also has limitations: it requires multiple data sets that cannot easily be merged, it relies at times on outdated data, and it requires complex methodological approaches to analysis.[8]

An alternative approach to data collection on pandemic response is to examine online information-seeking behavior through Google Trends (GT), a source of data on trends in online information-seeking.[9] Google Trends data has been leveraged extensively in health behavior research by using trends in online search queries as proxies for changes in human behavior.[10] To date, Google Trends data has been used to analyze a wide array of health behavior topics, such as the relationship between media coverage and information seeking around major infection events, the impact of local abortion policies on information seeking about abortion, the influence of influenza concerns on travel, and temporal patterns in interests in healthy behaviors such as dieting.[11-14] Initial research on COVID-19 in the US using Google Trends data has shown increases in general virus information seeking, but to date most studies explore a singular issue over time or in a specific geography, without examining comprehensive search pattern correlations or geographical heterogeneity, and their potential implications.[15-17] Google Trends offers immense potential for generating temporally and geographically specific insights into the beliefs, behaviors, and actions of various communities in their response to the coronavirus.

Here, we created a curated list of search queries to analyze changes in information seeking related to the COVID-19 pandemic, over time and across geographies in the US. *We answer four broad questions regarding information seeking around coronavirus:* **First**, how is information-seeking changing over time? Specifically, what is the relationship between the coronavirus outbreak and internet searches related to healthcare seeking, government support programs, media sources of different ideologies, planning around social activities, travel, and food, and new coronavirus-specific behaviors and concerns? We would expect certain types of searches (e.g. healthcare related inquiries, government safety net programs, online food delivery) to become more popular, while others fall in popularity (e.g. nearby bars, travel plans). These time trends provide clear contextual information on the changing interests and concerns of Americans as the pandemic progresses, and we can also compare the timing of these changes with real-world actions and events like policy announcements.

**Second**, what variation exists in information-seeking behavior at the regional and state levels? Specifically, how does the popularity of search terms differ across states and regions? Given the immense heterogeneity across the US, both culturally and in terms of COVID-19 caseload and response, we would expect states to differ in the information most sought out. Observed geographic differences in information seeking add granularity to the information-seeking context and provide an opportunity to develop hypotheses around why these differences exist.

**Third**, do states have particular and distinct patterns in information seeking? Specifically, what searches correlate with each other at the state level? Using machine learning methods, specifically unsupervised learning, we can construct different typologies of geographic areas, so that those whose search trends provoke concern for public health can be better targeted with information and other interventions.

**Fourth**, does Google Trends data correlate with and potentially precede real-life events? If Google Trends data can be “predictive” of real-life events (e.g. unemployment rates), it provides further validation for the method as a window into the behaviors of Americans during the pandemic.

To answer the above questions, queries were grouped into six themes relevant to COVID-19 behavior change: Social & Travel, Care Seeking, Government Programs, Health Programs, News & Influence, and Outlook & Concerns. Our findings demonstrate there have been both significant geographic heterogeneity and shifts over time in the information-seeking occurring across the six themes. Notably, there is high search interest in coronavirus symptoms and behaviors but declining interest in other health issues. The news media is critical to information seeking, and during the month of March 2020, Americans made some big shifts in information-seeking indicative of socio-behavioral changes even before government policy began. However, we also observed the relative increase in popularity of searches on financial concerns. While there is geographical variation in information-seeking patterns, we were also able to identify search patterns for specific states, pointing to potential public health vulnerabilities that should be addressed.

This study and analytic approach provide insights into, and potential indicators of, important changes in information-seeking patterns during pandemics like COVID-19. We propose specific strategic shifts for policy makers to improve the precision and effectiveness of non-pharmaceutical interventions (NPIs), and we also provide suggestions for developing a real-time dashboard as a decision-making tool - ongoing tracking of Google Trends could provide value in the coming phases of the epidemic as the states moves to the reopen their economies.

# Methods

Google Trends is a free online source of data on trends in online information-seeking across geographies (country, region, city, and designated market area/metro where this exists) and over time (since 2004). Historically, research using Google Trends has suffered from inconsistent methodology, poor documentation, and low reproducibility.[3] In 2019, a comprehensive methodological framework was published to standardize approaches to Google Trends research.[18] This work follows the framework, which provides specific criteria for selecting the keywords, geographical regions, and time period for analysis. We also combine several keywords that represent a similar topic into a single “query”.

## Keywords and Search Queries

Google Trends accepts a single word or a phrase, and several of these can be combined into a single Trends query by joining with a "+", which functions like an "OR". This study used 39 queries of up to 5 words/phrases each **[see list of queries in the appendix]**. We generated and curated search terms and evaluated each query's relevance and data quality in representing different aspects of the response to and impact of coronavirus search activity on Google Trends. Each query was categorized into one of 6 emergent themes: *Social & Travel, Care seeking, Government Programs, Health Programs, News & Influence, Outlook & Concerns* **[see list of queries in the appendix]**. For the News & Influence theme, we selected media outlets representative of both sides of the ideological spectrum based on a Pew Research survey conducted in 2014, measuring the ideological leanings (left or right) of the audience of different news media sources.[19]

## Data collection

Data from Google Trends (https://trends.google.com/) was extracted through an open-source 3rd party Python API, PyTrends. Based on its own documentation, Google Trends "*provides access to a largely unfiltered sample of actual search requests made to Google.*[20,21] *It is anonymized (no one is personally identified), categorized (determining the topic for a search query) and aggregated (grouped together). This allows us to display interest in a particular topic from around the globe or down to city-level geography.*" Google provides the relative search values (RSVs) for each query on a scale from 0 to 100, representing a normalized value, as explained in Google's documentation: "*Each data point is divided by the total searches of the geography and time range it represents to compare relative popularity. Otherwise, places with the most search volume would always be ranked highest. The resulting numbers are then scaled on a range of 0 to 100 based on a topic's proportion to all searches on all topics*."

Three primary datasets were developed from Google Trends: one at national level and the other two at state level, capturing all 50 states and the District of Columbia.

The national data set contained weekly RSVs for the US for each of 39 queries for the 224 weeks between Jan 1st, 2016 to April 15th, 2020. This allowed us to conduct trend analysis for each query independently, tracking its relative popularity in the entire US compared with its most popular week over these 224 weeks. For each query, the week when the query was most popular (as a percentage of all queries in that week) was scored as 100 by Google Trends, and all other weeks were scored relative to this week.

One of the state-level data sets contained weekly RSVs for each query for each state for the 16 weeks between January 1st, 2020 and April 15th 2020 while the other contained aggregated RSVs for each query for each state over the whole period of March 1st, 2020 to April 15th, 2020. For each of these data sets, RSVs for a specific query were always expressed as search popularity relative to other weeks, other states, or both; but never relative to other search queries (i.e. our approach factored out absolute differences in popularity between queries).

Finally, two additional, secondary data sets were referenced, comparing Google Trends against new monthly Medicaid applications and weekly initial unemployment claims, from https://data.medicaid.gov (June 2017 to Jan 2020) and https://oui.doleta.gov/unemploy/claims.asp (Jan 2016 to April 2020), respectively.

## Analysis

Geographically, states were grouped by *federal region and division* based on US census data.[22]

We used Python for data processing and Tableau for exploratory data analysis and visualization.[23] We used Python and R packages to perform pairwise correlation (Pearson, pairwise complete) and unsupervised learning - principal component analysis (PCA), respectively.[24] We conducted pairwise correlation to visualize and quantify associations between the individual queries at state level; we then applied dimensionality reduction through PCA to reveal groups of queries often searched together and identify top states with search patterns of interest - either supporting or undermining the fight against the coronavirus epidemic. Missing values were removed for correlation and PCA analysis, i.e. we kept all 51 geographies and removed any queries that had missing values for any states. Nine queries had missing values: *"coronavirus infowars + coronavirus breitbart + coronavirus glenn beck + coronavirus the blaze", "how can I stop coronavirus", "coronavirus can I see a doctor + coronavirus can I get a test + coronavirus are tests available", "coronavirus afford doctor + coronavirus uninsured + coronavirus medical bill", "bar closed + restaurant closed", "government aid", "doctor appointment", "doctor open + doctor office open", and "can't pay rent + how pay rent + behind on rent + can't pay mortgage"*. This reduced the number of queries for analysis from 39 to 30.

For pairwise correlation, we used a scale of 0-0.4 as low, 0.4-0.6 as moderate and 0.6-1.0 as high. We then focused only on the most prominent (moderate and high) scales. For PCA, only loadings with loading scores higher than 0.2 or lower than -0.2 are used to explain each component (PCA loadings take values from 0 to 1).

To calculate the increase in search popularity between January and March 2020 we took the mean RSV for March 2020 and divided this by the mean RSV for January 2020 and expressed this ratio as a percentage change. We chose January and March to capture the biggest shifts in RSV based on the progression of the virus. Data for the extra two weeks in April was used to monitor which trends maintained or dissipated after big shifts in March.

Finally, to compare Google Trends to real-life phenomena, we visualized data on actual weekly initial unemployment claims and new applications for Medicaid, then quantified the correlations between each phenomenon and the corresponding Google Trends query.

# Results

## National trends show changes in coronavirus information seeking related to symptoms, social distancing and other health and lifestyle activities

Analysis of all queries showed substantial shifts in RSVs across all thematic categories **[Fig 1]** predominantly in March 2020, with trends persisting or stabilizing in early April.

For Care Seeking, the RSVs of coronavirus-related queries increased substantially ("coronavirus symptoms" and "testing centers") while the RSVs of general health-seeking queries ("urgent care" and "doctor appointments") declined in late March/early April, suggesting a nuanced story of care seeking **[Fig 1].** This trend was matched by a decline in the RSVs for all queries in the Health Programs theme ("health insurance", "medicaid", and "medicare"), by 18%, 23%, and 26% respectively. Comparison with the same time window in preceding years confirmed this was an anomaly against historical seasonality, where health program searches peak and drop in November/December, not in March/April.

News & Influence searches related to coronavirus saw notable hikes in RSV for both left-leaning and right-leaning media. RSV for far-right/alt-right media outlets (see methods section for definitions based on Pew Research survey) also increased. Simultaneously, RSV for "coronavirus fake news" and "coronavirus hoaxes" surged 38-fold **[Fig 1]**.

For Outlook & Concerns, RSVs for new behavioral concepts gained immense popularity: "social distancing" and "how to make masks" spiked by 100-fold and peaked in March as did time-sensitive concerns like "hoarding" and "can't pay rent". In contrast, searches for "stock outs/sold out" and "coronavirus medical bill/affordability" remained at high levels into early April, hinting at potential differences in long-term versus short-term concerns **[Fig 1]**.

Within the Social & Travel theme, search trends were aligned with the new norm of social distancing and also signaled potential drops in business for the travel and service industries. RSVs for "online groceries + food delivery" and "food delivery + grocery deliveries + takeout + curbside + online food order" tripled and doubled respectively, while RSVs for "party ideas", "nearby bars and restaurants", and "cheap flights/travel" all dropped by between 17% and 51%. This shift is noteworthy for "nearby bars and restaurants", as it is a reversal in an upward trajectory which had recently peaked in December 2019 **[Fig 1]**

For Government Programs, relative popularity increased for new coronavirus-specific packages ("stimulus check" and "small business loans"), as well for unemployment benefits, a more mature program. Notably, the RSVs for "disability/food stamps" and "government aid" had the lowest increases in RSVs among government programs. However, relative interest in programs focused on individuals - “stimulus check”, and “unemployment application” - maintained momentum through April 15th [Fig 1]

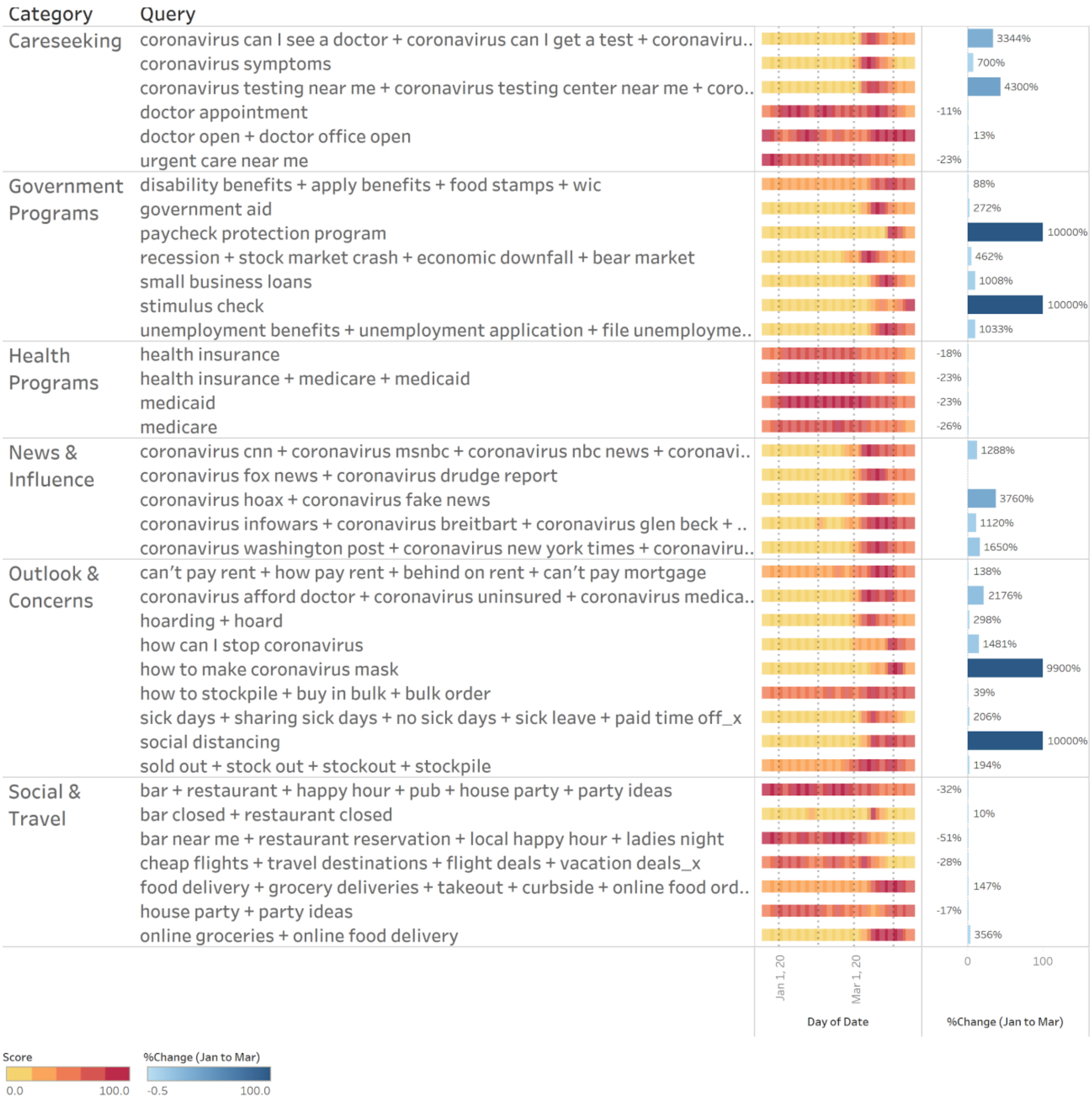

**Fig 1.** National search pattern for each query in every theme - RSV for 16 weeks between January 1st and April 15th, 2020. The right panel shows a change of monthly average RSV between January and March, capped at 10,000% for queries with extremely high changes.

## Shifts in information seeking occurred up to 12 days before federal government policy action

At the national level, Google Trends show that the relative popularity of queries related to non-pharmaceutical interventions (NPIs) was already shifting substantially, and in some cases peaking, several days ahead of major federal government policies and NPI action [Fig 2]. For example, RSVs for "social distancing" and "bar/restaurant nearby" were quickly increasing and decreasing, respectively, by March 8th - 3 days before the World Health Organization declared COVID-19 a pandemic, 5 days before the US federal government's national emergency declaration, and 8 days before the government released official social distancing guidelines. Furthermore, "how to make a coronavirus mask" was already relatively popular (RSV> 50) by March 23, exactly 12 days before the CDC advisory promoting masks for the general public.

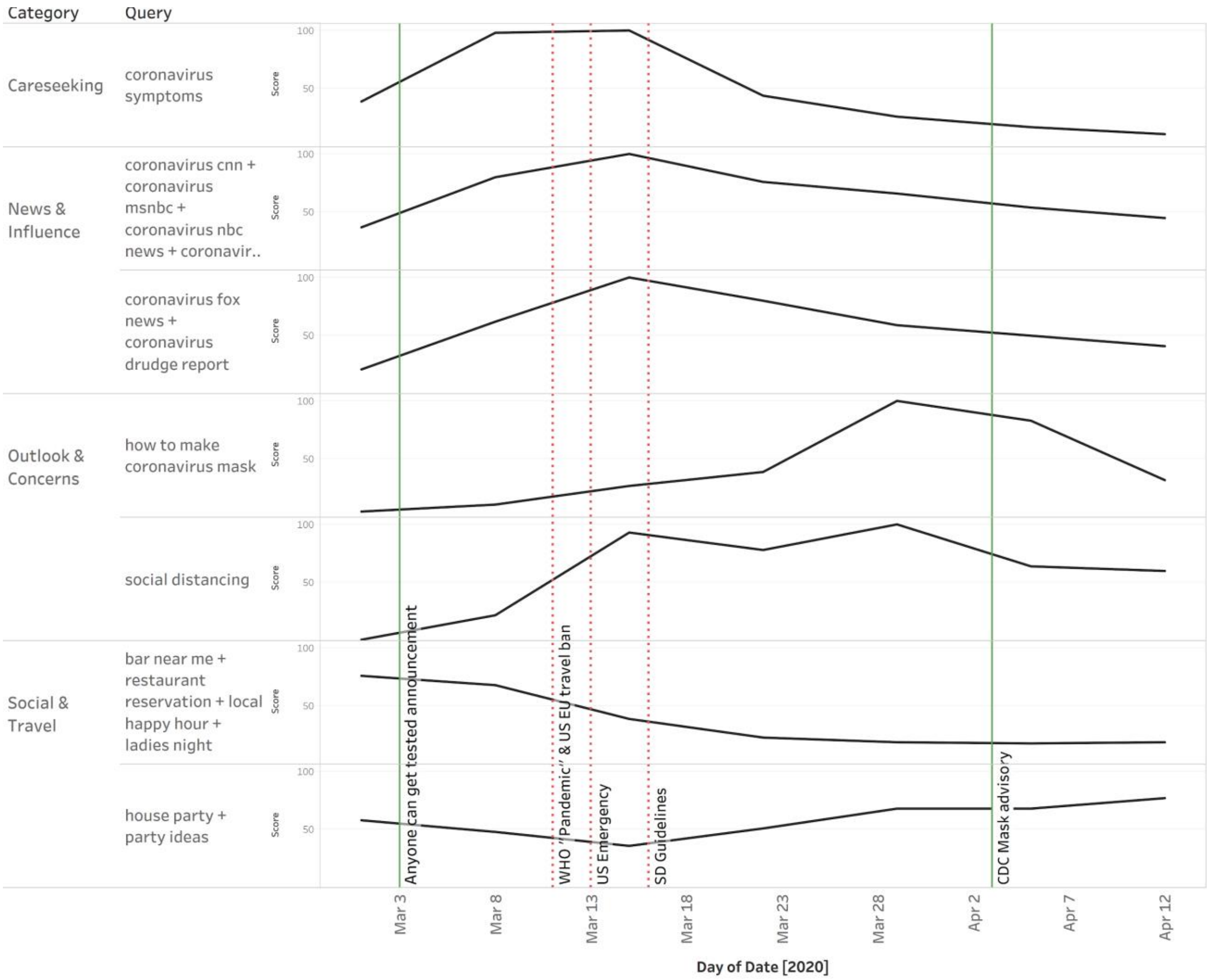

**Fig 2.** National trends for Google Trends NPI-related queries compared to actual government and public health NPIs at US federal level between March 1, 2020 and April 15th, 2020.

On the other hand, the RSV trend for news on coronavirus and RSV trend for coronavirus symptoms were in sync, matching each other regardless of the ideological leaning of the news source. This suggests that searches for disease characteristics were highly correlated with searches for news coverage of the disease.

## Google Trends also preceded important real-life events - weekly unemployment claims and new Medicaid monthly applications

Google Trends also showed potential value as a sentinel indicator for tracking and anticipating big shifts before they happen. We found high correlation between specific Google Trends queries and corresponding phenomena in the real world both before the epidemic (for unemployment and Medicaid) and during the epidemic (for unemployment). Two examples were selected - initial weekly unemployment claims and new monthly Medicaid applications. **Fig 3** shows the comparative results with corresponding Google Trends queries over time.

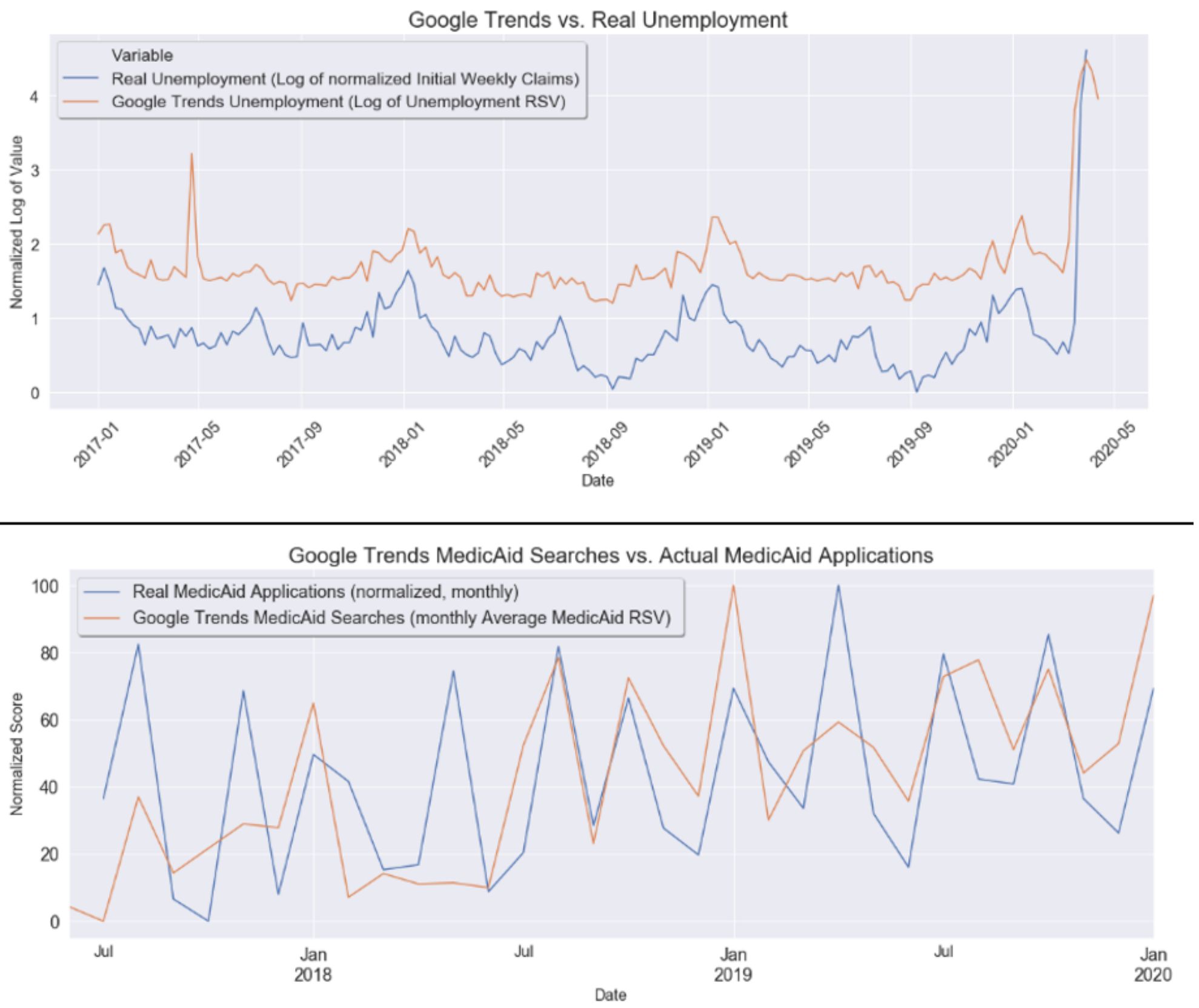

Fig 3. National-level multivariate trends for new weekly unemployment (Jan 2016 to April) and new monthly Medicaid applications (Jul 2017 to Jan 2020): **Top**:  Google searches for unemployment applications and actual initial weekly unemployment claims normalized from 0-100 over time (with seasonal adjustment). **Bottom:** Monthly average Google searches for Medicaid and actual new applications for Medicaid normalized from 0-100 (lagged by 1 month).

There was an extremely high correlation of **0.96** between the previous week's Google Trends RSV for unemployment applications and the actual weekly initial unemployment claims normalized to 0-100 (seasonally adjusted). For Medicaid, there was a high correlation of **0.55** between average monthly Google Trends RSV for Medicaid, lagged one month and the number of new applications for Medicaid each month.

However, we noted that this approach was only modestly successful for more complex cases where the link between search popularity and actual behavior is more difficult to imagine. A notable example is between stock market prices (Dow Jones Industrial Average) and the Google Trends query for "recession/stock market crash" which showed no correlation.

## Geographical heterogeneity of relative query popularity suggests varying responses to coronavirus across states and regions

Nationally, the biggest jumps in RSV were concentrated between March 1st and April 15th, so we focused on this window to investigate state-by-state differences in information seeking. Here, we compared the relative popularity of a search query across states in a single time period rather than across time.

Findings showed similar levels of popularity of searches for "urgent care near me" in the South and Northeast but a difference in relative popularity of searches for health programs (health insurance, Medicare, and Medicaid). Care Seeking searches were most popular in the South (Louisiana, Georgia, and North Carolina); the Northeast (New York and New Jersey); and in Indiana, Illinois and Arizona during this specific window **[Fig 4a]**. While New York and New Jersey were already COVID-19 hotpots, this period also coincided with big jumps in the number of cases for Louisiana, Georgia, and North Carolina.

RSVs for "social distancing" were generally high in the Northeast and West, but generally low in the South, while searches for "hoarding" were relatively more popular in Alaska, New Mexico, Minnesota, Arizona, and Montana than in other states during this period.

Regarding News & Influence, searches for far-right/alt-right and coronavirus were most popular in West Virginia, Oklahoma, Idaho, and Pennsylvania, while "fake news coronavirus" searches were most popular in DC, Vermont, Alaska, Maine, and Nebraska **[Fig 4b.].** Additionally, "stimulus check" had particularly high RSVs in Southern states **[Fig 4b].** While "health insurance" had a high RSV in the Northeast (New York, Massachusetts, Vermont), "Medicaid" had higher RSVs in Southern states, including Louisiana and Mississippi **[Fig 4b.]**.

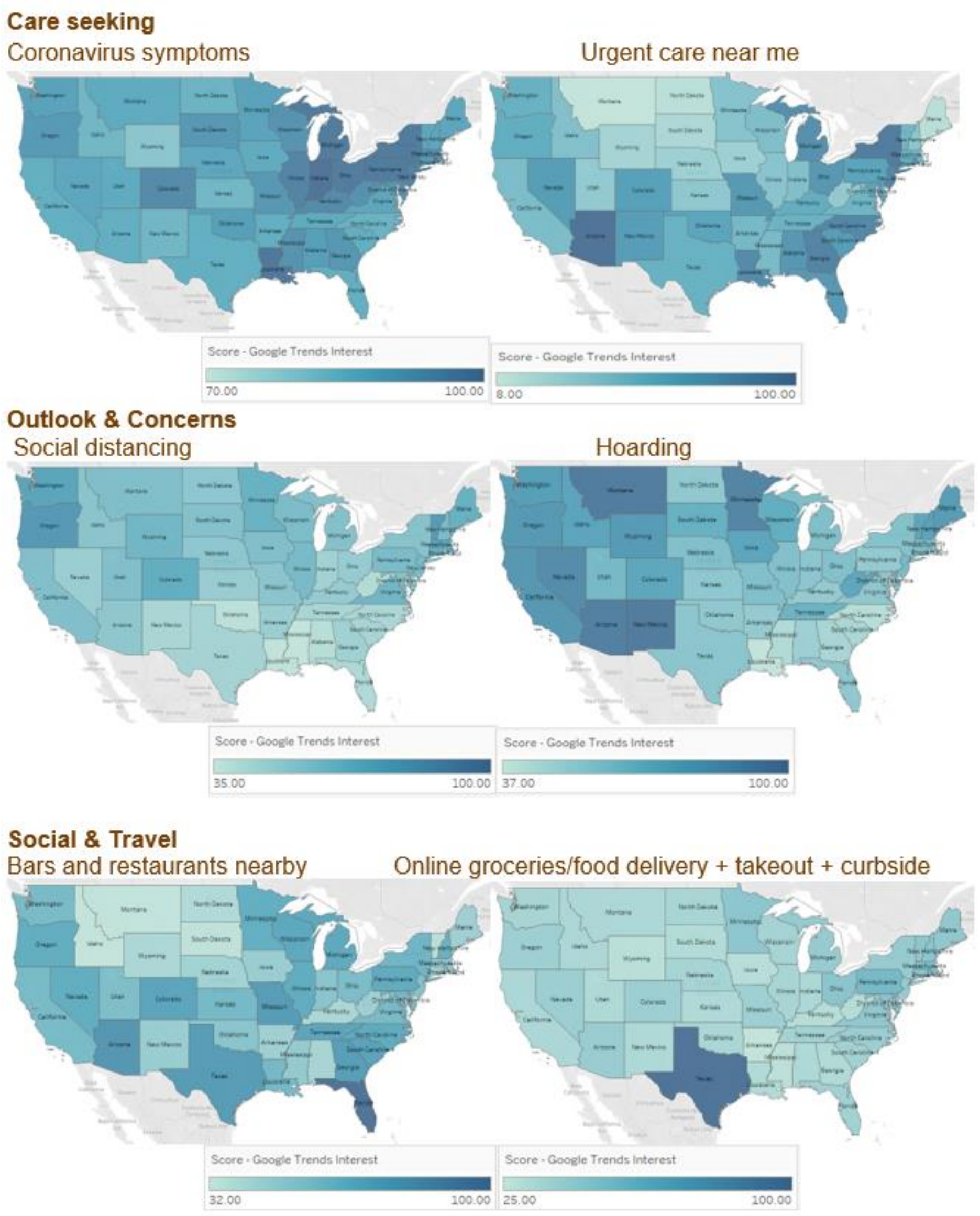

Fig 4a. Geographical variations across states, March 1st, 2020 to April 15th, 2020. Shown are the two queries most representative of each category ***[Care Seeking, Outlook & Concerns, Social & Travel]***

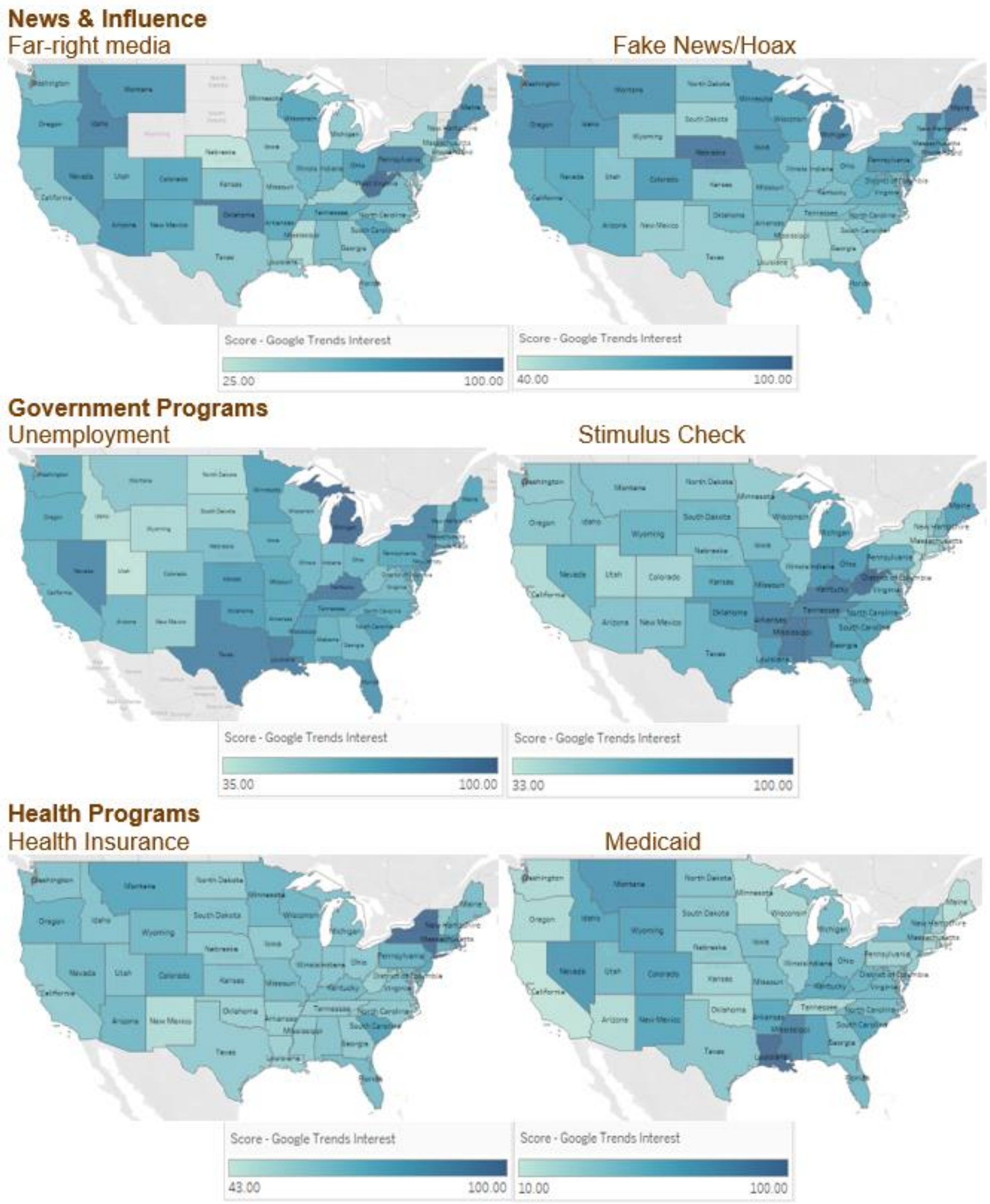

Fig 4b. Geographical variations across states - March 1st, 2020 to April 15th, 2020. Shown are the two queries most representative of each category ***[News & Influence, Government Programs, Health Programs]***

## A few components explain a lot of variability in information-seeking between states, identifying states with potentially problematic search patterns

While the thematic categories in the above analysis give insights on the relative popularity of queries over time and across geographies, we also conducted pairwise correlation and PCA analyses to understand how the queries were correlated. First, correlation analysis showed that states with high RSVs for “coronavirus symptoms” also tended to have high RSVs for “urgent care” and for “test centers”, suggesting a relationship between awareness of the disease and potential intention to seek care. Second, the correlation also showed that states with high RSVs for “social distancing” also tended to have high RSVs for coronavirus  and all types of news media regardless of ideological leaning, “recession/stock market crash”, and “sick days/leave” while tending to have low RSVs for safety net programs - “medicare” and “stimulus check”. This suggests that news sources and economic factors may play a role in the levels of interest in, awareness of, and potential adoption of social distancing behavior.

The results of PCA are shown in **Fig 5**, with all 39 queries summarized as the top 2 components, which explain **43%** of the variation (differences or patterns) in the data - 25% from Component 1 and 18% from Component 2. The two components were assigned labels based on search patterns they showed.

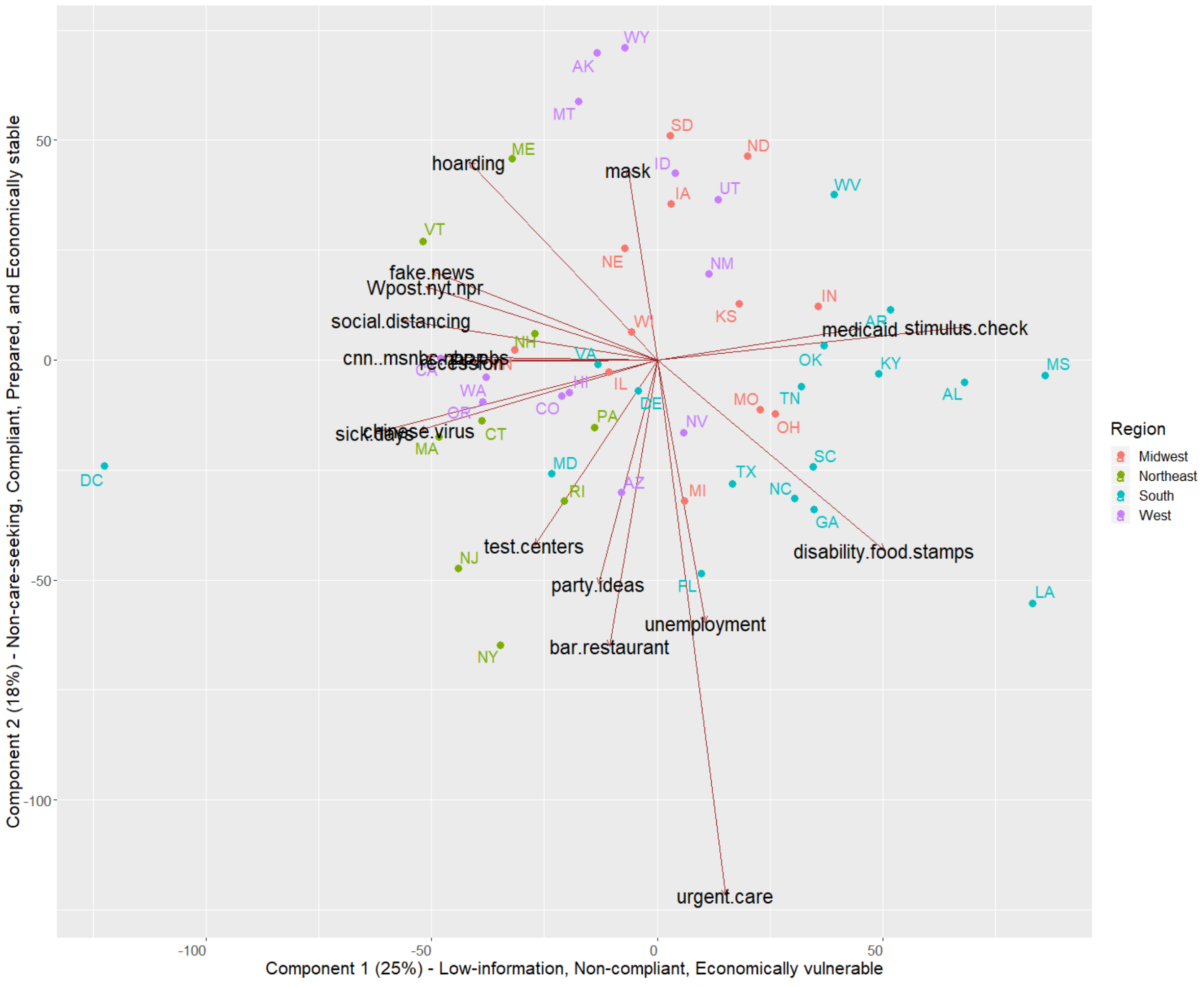

Fig 5. Scatterplot of state PCA loadings/scores for first 2 components with top queries shown as arrow vectors. Each arrow represents the relative weight each query has, and the direction indicates the points to the states most exhibiting this search pattern. Arrows direction measures correlation - if in the same direction are highly positively correlated while divergent arrows in opposite directions are highly negatively correlated. Component 1 (x-axis) and Component 2 (y-axis) explain 25% and 18% of variation (differences or patterns) in the data, respectively.

PCA revealed search patterns related to economic vulnerability and searching for information from news/media sources. It also highlighted how state search patterns were related to other concepts such as compliance with social distancing or related policies like mask wearing; preparation for emergencies (hoarding), and care-seeking concepts such as searching for urgent care.

Specifically, Component 1 explained 25% of the variation in the data **Fig 5**. This component represented potentially *Low-information, Non-compliant, Economically vulnerable* states. These terms are defined as follows: *Low-information* - low association with searches for any news source, whether real or fake; *Non-compliant* - low association with searches for "social distancing"; and *Economically vulnerable* - high association with searches for "disability/food stamps" and "stimulus check". The states with the highest score, i.e. those that exhibited this search pattern most strongly, were Mississippi, Louisiana, Alabama, Arkansas, and Kentucky. There are also additional clusters of states within Component 1 that have high scores for "urgent care nearby" and "unemployment application" - Florida and Michigan - as well as "disability/food stamps" - Georgia, North, and South Carolina.

Component 2 explained 18% of variation in the data **Fig 5**. This represented potentially *Non-care-seeking, Compliant, Prepared, and Economically stable* states. These terms are defined as follows: *Non-care-seeking* - low association with searches for urgent care nearby; *Compliant* - low association with searches of nearby bars/restaurants; *Prepared* - high association with searches for "how to make a mask" and "hoarding"; and *Economically stable* - low association with searches for "unemployment" and "disability/food stamps". The states exhibiting this trend were Wyoming, Alaska, Montana, as well as North and South Dakota.

# Discussion

This study shows substantial changes in coronavirus information-seeking at the national level, particularly in March 2020, suggesting a hyper-awareness and desire for information about both coronavirus and its corresponding novel behavioral concepts, i.e., social distancing and mask wearing. Trends also mirrored the rapid changes in the way people eat, travel, and socialize, with increased interest in online food/groceries and rapidly declining interest in bars/restaurants and cheap flights. Searches for hoarding or stockpiling also increased, suggesting preparation for social distancing or emergencies. However, there was also substantial increased search popularity around ability to pay rent and/or coronavirus medical bills as well as availability of government benefits to weather the crisis. This indicated that some could not afford to shift their lifestyle to appropriately protect themselves against the pandemic, a key insight for policymakers into American's concerns.

The high demand for information also corresponded with increasing searches for news sources and coronavirus, regardless of ideological leaning, including searches for "coronavirus fake news + coronavirus hoax". RSVs for coronavirus and ideologically left-leaning news sources were highly correlated with RSVs for fake news, but it is not possible to conclude whether this indicates curiosity or an earnest belief in the existence of fake news. Regardless, this finding underscores the critical and timely role that news sources play in providing information during a pandemic and why this information must be correct and trusted.

In contrast, there was almost no discernible temporal relationship between national government policy/NPIs and the search popularity of corresponding ideas. In fact, increases in information-seeking for social distancing and drops in search for nearby bars/restaurants were already significantly declining a full 8 days before the government released official social distancing guidelines. Similarly, Americans were already searching for how to make masks 12 days before the CDC recommended wearing them. This suggests that while there are political, logistical, and other considerations, Google Trends can provide an indication of new and emerging behavioral issues to inform public health policy, by showing what people are already aware of and what interventions may be acceptable. Monitoring Google Trends can provide insights into an optimal approach to action, especially for fast-evolving situations. The right interventions, poorly timed, could be too early to be acceptable, or, as occurred with the coronavirus pandemic case, too late to be optimally effective. This also highlights the importance of local authorities in the fight against coronavirus. It is likely that shifts in search interests were more closely related to early local government announcements that preceded the national and global announcements used in our analysis.

As the relative popularity of coronavirus care-seeking searches increased, search interest in other health behaviors (urgent care, doctors' appointments, health insurance/Medicare/Medicaid) decreased. This is consistent with the decline in health seeking that occurs during pandemics: prospective patients for other diseases have higher risk perception of hospital-based transmission of COVID-19, which reduces their health-seeking behavior.[25] This is especially true in the US, where providers were instructed to prioritize coronavirus and relegate other services.[26] This finding has potential value as an early-warning sign of reduced health-seeking behavior. It could be of particular use in especially vulnerable communities with underlying health risks, allowing authorities to prioritize and precisely target these areas with messages about the continued importance of health care for non-coronavirus related conditions and symptoms.

While social distancing queries had a similar popularity across most states, most queries showed geographical heterogeneity in their relative popularity. This provides more granular context and an opportunity to propose plausible hypotheses to explain some of the differences. These hypotheses would need to be validated, but they provide a starting point for anticipating the trajectory of the next pandemic or national emergency.

State-by-state heterogeneity for health-seeking queries confirmed both the trajectory of the pandemic and potential regional and structural drivers of these differences. For example, in New York and New Jersey, two states with a more mature pandemic and, coincidentally, state-run (rather than federally run) health insurance marketplaces, people were searching for urgent care and for health insurance. Southern states tended to search for several potentially worrying factors in the fight against coronavirus, including searches for urgent care as the virus arrived (Louisiana, Georgia, and North Carolina), searching for Medicaid (Louisiana and Mississippi), and searches for  stimulus checks indicating a needed financial safety net (most Southern states). The search for “hoarding” was popular in states with either large land areas (and less dense populations) or with especially extreme weather conditions - Alaska, New Mexico, Minnesota, Arizona, and Montana. This may reflect their populations’ heightened expectations of scarcity during a pandemic, as these factors likely make these states hard to reach in the case of disrupted supply chains. However, it was not clear why Washington, D.C. had several queries, both positive and negative, as top concerns during March and April 2020. One hypothesis is that this is a product of D.C.’s role as the national capital, home to prominent government and news agencies providing and consuming coronavirus information.

We also systematically captured and described the heterogeneity by extracting search patterns using correlation analysis and PCA to condense dozens of queries to two simple components, with actionable implications for identified target geographies if concerning patterns emerge. While correlation analysis and PCA are not new methods, this study expands the methodological approaches by applying both to a unique dataset. When combined, richer and more specific insights emerge - with both analyses identifying economic/financial factors, access to information, and/or interest in social distancing as key variables in describing states' information-seeking patterns on coronavirus. With PCA we identify Mississippi, Louisiana, Alabama, Arkansas, and Kentucky as *Low-Information*, *Non-Compliant*, and *Economically Vulnerable* states during the time window of analysis March/April 2020. This search pattern suggests that the states most vulnerable economically or in terms of the social safety net are also the least informed and show the least search-related interest in social distancing, a key intervention against coronavirus. This information can be combined with real-world data to increase awareness of social distancing and its benefits, while targeting and prioritizing resources to support these states - including increased testing, health system capacity supports, and economic relief measures for the population.

To capture the evolving information and insights available through Google Trends we propose a real-time dashboard to track trends, geographical variations, and patterns of interest regarding the epidemic.  The queries used in this study can be used as a starting point or baseline with additional features added if needed. As the epidemic continues and potentially gives way to a second wave, this dashboard will follow how specific queries change over time, and in which states we see concerning search patterns using the pairwise correlation and PCA approaches. Depending on the search patterns identified, policymakers can then design or improve interventions, as well as allocate resources to target states.

We made some assumptions and choices which may result in limitations depending on the use case. First, selection of keywords and queries was an iterative process because it is difficult to a priori know what search terms best capture a desired concept. Analytically, we worked at the higher levels of aggregation to capture the most comprehensive units of analysis. When investigating trends over long periods of time, beyond one month, we used weekly, as opposed to daily, RSVs. We also chose to work with the state as our main unit of analysis. We found that analysis at more granular Designated Market Area (DMA) or city level as provided by Google Trends was not robust enough for our purposes, because the coronavirus response is organized and determined at the state level in many cases. Furthermore, many DMAs span multiple states which introduces complicating effects, and the city level often has data missing, leaving the state level as the best option. Even at state level, there were a few missing values which were included in the trend and geographical analyses but removed for pairwise correlation and PCA analysis.

While valuable insights can be derived, a major limitation of Google Trends is that it will always be a measure of search patterns, and not the actual corresponding behaviors, so inference and prediction must always come with this caveat. Finally, this analysis is limited by Google Trends' specific assumptions, which include pulling the data from only a sample of and not the whole database of searches, and providing it in the form of a relative search value (RSV) instead of absolute search volumes per geography. This requires caution when analyzing results from low-volume queries or geographies and when making interpretations and conclusions from analyses.

## APPENDIX

Final list of search queries grouped into emergent themes

| **Category/ Theme** | **Query [39 queries with 1 to 5 combined terms]** |
|---|---|
| ***Care Seeking*** | *coronavirus symptoms*<br>*coronavirus testing near me + coronavirus testing center near me + coronavirus test*<br>*doctor appointment*<br>*coronavirus afford doctor + coronavirus uninsured + coronavirus medical bill*<br>*Coronavirus can i see a doctor + coronavirus can i get a test + coronavirus are tests available*<br>*doctor open + doctor office open*<br>*urgent care near me* |
| ***Government Programs*** | *disability benefits + apply benefits + food stamps + wic*<br>*government aid*<br>*recession + stock market crash + economic downfall + bear market*<br>*small business loans*<br>*stimulus check*<br>*paycheck protection program* |
| ***Health Programs*** | *health insurance*<br>*health insurance + medicare + medicaid*<br>*medicaid*<br>*medicare* |
| ***News & Influence*** | *chinese virus*<br>*coronavirus cnn + coronavirus msnbc + coronavirus nbc news + coronavirus cbs news*<br>*coronavirus hoax + coronavirus fake news*<br>*coronavirus fox news + coronavirus drudge report*<br>*coronavirus infowars + coronavirus breitbart + coronavirus glenn beck + coronavirus the blaze*<br>*coronavirus washington post + coronavirus new york times + coronavirus npr*<br>*coronavirus fox news + coronavirus drudge report* |
| ***Outlook & Concerns*** | *can't pay rent + how pay rent + behind on rent + can't pay mortgage*<br>*hoarding + hoard*<br>*how can i stop coronavirus*<br>*how to make coronavirus mask*<br>*how to stockpile + buy in bulk + bulk order*<br>*sick days + sharing sick days + no sick days + sick leave + paid time off*<br>*social distancing*<br>*sold out + stock out + stockout + stockpile*<br>*unemployment benefits + unemployment application + file unemployment + apply unemployment + layoffs* |
| ***Social & Travel*** | *bar closed + restaurant closed*<br>*bar near me + restaurant reservation + local happy hour + ladies' night*<br>*bar + restaurant r + happy hour + pub + house party + party ideas*<br>*cheap flights + travel destinations + flight deals + vacation deals*<br>*food delivery + grocery deliveries + takeout + curbside + online food order*<br>*house party + party ideas* |

## References

1. *Survey Tool and Guidance: Rapid, simple, flexible behavioural insights on COVID-19.* Copenhagen: WHO Regional Office for Europe 2020.
2. Rubin GJ, Amlôt R, Page L, Wessely S. Methodological challenges in assessing general population reactions in the immediate aftermath of a terrorist attack. *International Journal of Methods in Psychiatric Research.* 2008;17(S2):S29-S35.
3. Nuti SV, Wayda B, Ranasinghe I, et al. The Use of Google Trends in Health Care Research: A Systematic Review. *PloS one.* 2014;9(10):e109583.
4. Rubin GJ, Bakhshi S, Amlot R, Fear N, Potts HWW, Michie S. *The design of a survey questionnaire to measure perceptions and behaviour during an influenza pandemic: the Flu TElephone Survey Template (FluTEST).* Southampton (UK): Health Services and Delivery Research; 2014.
5. Nelson LM, Simard JF, Oluyomi A, et al. US Public Concerns About the COVID-19 Pandemic From Results of a Survey Given via Social Media. *JAMA Internal Medicine.* 2020.
6. Ashley Kirzinger AK, Liz Hamel, Mollyann Brodie. KFF Health Tracking Poll - Early April 2020: The Impact Of Coronavirus On Life In America. Keiser Family Foundation. https://www.kff.org/health-reform/report/kff-health-tracking-poll-early-april-2020/. Published 2020. Accessed.
7. Salathé M. Digital Pharmacovigilance and Disease Surveillance: Combining Traditional and Big-Data Systems for Better Public Health. *J Infect Dis.* 2016;214(suppl_4):S399-S403.
8. Buckee C. Improving epidemic surveillance and response: big data is dead, long live big data. *The Lancet Digital Health.* 2020.
9. Google Trends. Google. https://trends.google.com/trends/?geo=US. Accessed.

10. Kan W-C, Chien S, Wang H-Y, Chou W. The most cited articles on the topic of health behaviors in Google Trends research: a systematic review. *Advances in General Practice of Medicine.* 2018:1-7.
11. Kostkova P, Fowler D, Wiseman S, Weinberg JR. Major infection events over 5 years: how is media coverage influencing online information needs of health care professionals and the public? *J Med Internet Res.* 2013;15(7):e107-e107.
12. Reis BY, Brownstein JS. Measuring the impact of health policies using Internet search patterns: the case of abortion. *BMC Public Health.* 2010;10:514-514.
13. Fenichel EP, Kuminoff NV, Chowell G. Skip the trip: air travelers' behavioral responses to pandemic influenza. *PloS one.* 2013;8(3):e58249-e58249.
14. Carr LJ, Dunsiger SI. Search query data to monitor interest in behavior change: application for public health. *PloS one.* 2012;7(10):e48158-e48158.
15. Benjamin Lucas BE, Todd Landman. Online Information Search During COVID-19. *arXiv.* 2020;preprint arXiv:2004.07183 [cs.CY].
16. Strzelecki A. The second worldwide wave of interest in coronavirus since the COVID-19 outbreaks in South Korea, Italy and Iran: A Google Trends study. *Brain Behav Immun.* 2020:S0889-1591(0820)30551-30551.
17. Hu DaL, Xiaoqi and Xu, Zhiwei and Meng, Nana and Xie, Qiaomei and Zhang, Man and Zou, Yanfeng and Liu, Jiatao and Sun, Guo-Ping and Wang, Fang. More Effective Strategies are Required to Strengthen Public Awareness of COVID-19: Evidence from Google Trends. *Available at SSRN: https://ssrncom/abstract=3550008.*
18. Mavragani A, Ochoa G. Google Trends in Infodemiology and Infoveillance: Methodology Framework. *JMIR Public Health Surveill.* 2019;5(2):e13439.
19. *Political Polarization & Media Habits.* Pew Research Center;2014.
20. Mills G. pytrends. GitHub. Accessed.
21. FAQ about Google Trends data. Google. https://support.google.com/trends/answer/4365533?hl=en. Accessed.
22. Census Regions and Divisions of the United States. In: U.S. Department of Commerce Economics and Statistics Administration U.S. Census Bureau.
23. Oliphant TE. Python for Scientific Computing. *Computing in Science & Engineering.* 2007;9(3):10-20.
24. Team RC. R: A language and environment for statistical computing. In. Vienna, Austria: R Foundation for Statistical Computing; 2019.
25. Ateev Mehrotra MC, David Linetsky, Hilary Hatch, and David Cutler. What Impact Has COVID-19 Had on Outpatient Visits? In. *To the Point (blog)*: Commonwealth Fund; 2020.
26. Healthcare Facilities: Preparing for Community Transmission. Center for Disease Control and Prevnetion. https://www.cdc.gov/coronavirus/2019-ncov/hcp/guidance-hcf.html?CDC_AA_refVal=https%3A%2F%2Fwww.cdc.gov%2Fcoronavirus%2F2019-ncov%2Fhealthcare-facilities%2Fguidance-hcf.html. Published 2020. Accessed.